\documentstyle[preprint,prb,aps,epsf,eqsecnum]{revtex}
\begin{document}
\draft

\title{Quantum-classical crossover in the spin 1/2 XXZ chain}

\author{Klaus Fabricius
\footnote{e-mail Klaus.Fabricius@theorie.physik.uni-wuppertal.de}}
\address{ Physics Department, University of Wuppertal, 
42097 Wuppertal, Germany}
\author{Barry~M.~McCoy
\footnote{e-mail mccoy@insti.physics.sunysb.edu}}               
\address{ Institute for Theoretical Physics, State University of New York,
 Stony Brook,  NY 11794-3840}
\date{\today}
\preprint{ITPSB-98-37}

\maketitle

\begin{abstract}
We compute, by means of exact diagonalization of systems of $N=16$ and
$18$ spins, 
the correlation function $<\sigma^z_0\sigma^z_n>$ at nonzero
temperature for the XXZ model with anisotropy $\Delta.$ In the
gapless attractive region $-1<\Delta<0$ for fixed
separation the temperature can always be made sufficiently 
low so that
the correlation is always negative for $n\neq 0$. However we find that for sufficiently large
temperatures and fixed separation or for fixed temperature greater 
than some $T_0(\Delta)$ and sufficiently large
separations the correlations are
always positive. This sign changing effect has not been previously seen and we
interpret it as a crossover from quantum to classical behavior.

\end{abstract}
\section{Introduction}

The spin 1/2 XXZ spin chain
\begin{equation}
H={1\over 2}\Sigma_{j=0}^{N-1} (\sigma_j^x\sigma_{j+1}^x+
\sigma_j^y\sigma_{j+1}^y+\Delta \sigma_j^z\sigma_{j+1}^z)
\label{ham}
\end{equation}
was first exactly studied by Bethe \cite{bet} in 1931 for $\Delta=\pm 1,$ 
extensively investigated for $\Delta=0$ in 1960 by Lieb, Schultz and Mattis
\cite{lsm}  and
studied for general values of $\Delta$ in 1966 by Yang and Yang
\cite{yya}-\cite{yyc}. Since these initial studies the thermodynamics have
been extensively investigated \cite{ts} and by now are fairly well understood.
The spin-spin correlations, however, are much more difficult to study
and even the simplest of the equal time
correlations
 \begin{equation}
S^z(n;T,\Delta)={\rm Tr}\sigma_0^z \sigma_n^z e^{-H/kT}/{\rm Tr}e^{-H/kT}
\label{sz}
\end{equation}
is only partially understood after decades of research
\cite{lsm,nm}-\cite{kss}.
In this note we extend these investigations of $S^z(n;T,\Delta)$  
for $T>0$ by means of an exact
computer diagonalization of chains of $N=16$ and $18$ spins 
for $-1\leq \Delta \leq 0.$ Our results are
presented below in tables 1-7 and Figs. 1 and 2. In the remainder of this note we
discuss the significance and the interpretation of these results. 

\section{Results and Discussion}

The correlation $S^z(n;T,0)$ for the case $\Delta=0$ was exactly
computed long ago \cite{lsm} to be
\begin{equation}
S^z(n;T,0)=\cases{-[{1\over \pi}\int_0^{\pi}d\phi
\sin(n\phi)\tanh({1\over kT}\sin\phi)]^2&if $n$ is odd\cr
\delta_{n,0}&if $n$ is even.}
\label{zerocor}
\end{equation}
This correlation is manifestly never positive for $n\neq 0.$ When $T=0$ it
simplifies to
\begin{equation}
S^z(n;0,0)=\cases{-4\pi^{-2}n^{-2}& if $n$ is odd\cr
\delta_{n,0}&if $n$ is even.}
\end{equation}
In the scaling limit where
\begin{equation}
T\rightarrow 0,~~n\rightarrow \infty,~~{\rm with}~~ Tn=r~~{\rm fixed}
\label{scaling}
\end{equation}
we have \cite{bm}
\begin{equation}
{\rm lim} T^{-2}S^z(n;T,0)=\cases{-\sinh^{-2}(\pi r/2)&if $n$ is odd\cr
0&if $n$ is even.}
\label{corr}
\end{equation}

In the general case $\Delta\neq 0$ the nearest neighbor 
correlation at $T=0$ $S^z(1;0,\Delta)$ is obtained
from the derivative of the ground state energy \cite{yyb} with respect
to $\Delta.$
This correlation is negative for $-1<\Delta$ and is plotted in Fig. 3
of  ref\cite{jkm}. 
For large $n$  the behavior of $S^z(n;0,\Delta)$ at $T=0$ has been
extensively investigated and for $|\Delta|<1$ we have \cite{lp,fog} for
$n\rightarrow \infty$
\begin{equation}
S^z(n,;0,\Delta)\sim -{1\over \pi^2 \theta n^2}+(-1)^n{C(\Delta)\over
n^{1\over \theta}}
\label{corra}
\end{equation}
where from ref.\cite{jkm}
\begin{equation}
\theta={1\over 2}+{1\over \pi}\arcsin \Delta.
\end{equation}
We note that $0\leq\theta\leq 1$ and vanishes at the ferromagnetic
point $\Delta=-1.$ At $\Delta=0$ we have
$\theta=1/2,~C(0)=2\pi^{-2}$ and
(\ref{corra}) reduces to the exact result (\ref{corr}). For other values
of $\Delta$ only the limiting value as $\Delta\rightarrow 1$ is 
known \cite{aff}. 

When $T>0$ the correlations decay exponentially for large $n$ instead
of the algebraic decay (\ref{corra}) of $T=0.$ For $0<\Delta<1$ it is
known \cite{klu,kbi} that for small fixed positive $T$ the large
$n$ behavior of $S^z(n;T,\Delta)$ is
\begin{equation}
S^z(n;T,\Delta)\sim
A_z(\Delta,T)(-1)^ne^{-nkT\pi(1-\theta)\theta^{-1}(1-\Delta^2)^{-1/2}}.
\label{corrb}
\end{equation}
In order to smoothly connect to the $T=0$ result (\ref{corra}) we need
$A_z(\Delta,T)=A(\Delta)T^{1/\theta}$ but this has not yet been demonstrated.
We note that for positive values of $\Delta$ the exact nearest
neighbor correlation at $T=0$
is negative and the leading term in the asymptotic behaviors (\ref{corra}) and
(\ref{corrb}) oscillates as $(-1)^n.$
Both of these facts are consistent with antiferromagnetism.

For negative values of $\Delta,$ however, the situation is somewhat
different.  The nearest neighbor correlation at $T=0$ 
is negative and, indeed,
since $\theta<1/2,$ we see from (\ref{corra}) that the asymptotic 
values of $S^z(n;0,\Delta)$ are
also negative and there are no oscillations. This behavior cannot be
called antiferromagnetic because there are no oscillations but neither
can it be called ferromagnetic because the correlations are negative
instead of positive.

In order to further investigate the regime $-1<\Delta<0$
we have computed the correlation function $S^z(n;T,\Delta)$  by means
of exact diagonalization for systems of $N=16$ and $N=18$ spins. Our
results for $N=18$ with $\Delta=-.1,-.3,-.9$ and $-1.0$ are given in
tables 1-4 where we give $S^z(n;T,\Delta)$ for $1\leq n \leq 8$ and
${1 \over 2}S^z(n;T,\Delta)$ for $n=9.$ The factor of $1/2$ for $n=9$ 
is used because for $n=N/2$ there are two paths of equal length joining $0$ and $n$
 in the 
finite system whereas for the same $n$ in the infinite size system 
there will be only one path of finite length.
To estimate the precision with which the $N=18$
system gives the $N=\infty$ correlations we give in table 5 the
correlation for $N=16$ and $\Delta=-0.9.$ 
We see here that for $T\geq .5$ 
the $N=18$ correlations are virtually identical with the
$N=16$ correlations. Even for
$T=.1$ and $T=.2$ the $N=18$ data should be qualitatively close to the $N
=\infty$ values.

The tables 1-5 reveal for $-1\leq \Delta \leq 0$ the striking 
property that $S^z(n;T,\Delta),$ which is
always negative at $T=0,$ becomes positive for fixed n at sufficiently large $T.$ 
We study this further in table 6 where we list the values $T_0(n;\Delta)$ where
$S^z(n;T_0(n;\Delta),\Delta)=0.$ This table indicates that
\begin{equation}{\rm lim}_{n\rightarrow \infty}T_0(n;\Delta)>0.
\end{equation}
We denote this limiting temperature by $T_0(\Delta)$ and 
note that this implies that in the expansion of 
$S^z(n;T,\Delta)$ obtained from the  
quantum transfer matrix formalism \cite{klu} 
\begin{equation}
S^z(n;T,\Delta)=\sum_{j=1}C_j(T;\Delta)e^{-n\gamma_j(T)}~~
{\rm with}~\gamma_j<\gamma_{j+1}
\end{equation}
we have $C_1(T_0(\Delta);\Delta)=0.$ If, for large $n,$ we retain only the 
first two terms in the expansion, ignore the $T$ 
dependence of $\gamma_j(T)$ and $C_2(T;\Delta)$ 
and write $C_1(T;\Delta)=(T-T_0(\Delta))C_1(\Delta)$
we see that
the large $n$ behavior of 
$T_0(n;\Delta)$ may be estimated as
\begin{equation}
T_0(n,\Delta)=T_0(\Delta)+Ae^{-n\gamma}.
\label{fit}
\end{equation}
where $\gamma=\gamma_2-\gamma_1$ and $A=-C_2/C_1.$
In Fig. 1 plot the data of table 6 
versus a least squares fit using (\ref{fit}) and find 
that the fit is exceedingly good even  for small $n.$
The values of the fitting parameters are given in table 7 for 
$-.9\leq \Delta \leq -.1$ and $T_0(\Delta)$  is plotted in Fig. 2. The 
existence of this $T_0(\Delta)>0$ for $-1<\Delta<0$ is quite 
different from the case $0<\Delta<1$ where for all temperatures 
the sign of $S^z(n;T,\Delta)$ is $(-1)^n.$

To interpret the property of changing sign we note that, when the
Hamiltonian (\ref{ham}) is written in  terms of the basis where
$\sigma^z_j$ is diagonal $(\sigma_j^z=\pm 1),$ the term 
$\sigma^x_j\sigma^x_{j+1}+ 
\sigma^y_{j}\sigma^y_{j+1}$ is a kinetic energy term which translates
a down spin one step whereas the term $\sigma^z_j\sigma^z_{j+1}$ is a
potential energy term which is diagonal in the basis of eigenstates of
$\sigma^z_j.$ In classical statistical mechanics the static 
expectation values of
position dependent operators are independent of the kinetic energy and
depend only on the potential energy. If  we further expect that at
high temperatures the system should behave in a classical fashion we
infer that at high temperatures for $\Delta<0$ the correlation
$S^z(n;T,\Delta)$ should be ferromagnetically aligned ie. $S^z(n;T,\Delta)>0.$
This is indeed what is seen in tables 1-5. However at low temperatures
the quantum effects of the kinetic term cannot be ignored. When $\Delta=0$
there is no potential energy so all the behavior in $S^z(n;T,0)$ can
only come from the kinetic terms and hence the behavior given by 
(\ref{zerocor}) in which $S^z(n;T,0)$ is never positive for $n\neq 0$ must be 
purely quantum mechanical. Consequently it seems appropriate to refer
to the change of sign of the correlation $S^z(n;T,\Delta)$ as a
quantum to classical crossover. 

The low temperature behavior of the correlation function is determined by conformal
field theory. In particular we consider the scaling limit (\ref{scaling}) and
define the scaling function
\begin{equation}
f(r,\Delta)=\lim T^{-2}S^z(n;T,\Delta).
\end{equation}
The prescription of conformal field theory is that this scaling
function is obtained from the large $n$ behavior of the $T=0$
correlation given by the first term of (\ref{corra}) 
by the replacement [page 513 of 
ref.\cite{kbi}]
\begin{equation}
n\rightarrow (\kappa T/2)^{-1}\sinh \kappa r/2
\label{replace}
\end{equation} 
where the decay constant $\kappa$ can be obtained by use of the
methods of ref. \cite{klu}.
This replacement is obtained by combining the conformal field 
theory results on
finite size corrections \cite{car} with the field theory relation of
finite strip size to nonzero temperature \cite{affb}.
This prescription clearly leads to a correlation which is
always negative and does not show the sign changing phenomena seen in
the tables 1-5. However, this result is only a limiting result as 
$T\rightarrow 0.$ The results of this paper indicate that there is further 
physics in the high temperature behavior of the XXZ chain where 
$T>T_0(\Delta)$ which is 
not contained in this conformal field theory result.

\acknowledgments

We are pleased to acknowledge useful discussions with  A. Kl{\"u}mper,
V. Korepin, S. Sachdev and J. Suzuki.
This work is supported in part by the National Science Foundation
under grant DMR 97-03543.

\begin{table}[b]
\begin{tabular}{|c|c|c|c|c|c|c|c|c|c|} 
$ T $ &$n=1$ & $n=2$ & $n=3$ & $n=4$ & $n=5$& $n=6$ & $n=7$& $n=8 $ & $n=9$ \\  
\hline
  0.1 & $-$3.81e-01 &  $-$1.87e-02 &  $-$3.74e-02 &  $-$4.51e-03 &  $-$1.19e-02 &  $-$2.01e-03 &  $-$5.81e-03 &  $-$1.28e-03 &  $-$2.21e-03 \\
  0.2 & $-$3.69e-01 &  $-$1.75e-02 &  $-$2.82e-02 &  $-$3.13e-03 &  $-$5.72e-03 &  $-$8.43e-04 &  $-$1.47e-03 &  $-$2.99e-04 &  $-$3.51e-04 \\
  0.5 & $-$2.78e-01 &  $-$1.10e-02 &  $-$5.88e-03 &  $-$4.65e-04 &  $-$2.27e-04 &  $-$2.24e-05 &  $-$9.31e-06 &  $-$1.11e-06 &  $-$3.86e-07 \\
  1.0 & $-$1.29e-01 &  $-$3.25e-03 &  $-$3.25e-04 &  $-$1.28e-05 &  $-$1.18e-06 &  $-$5.28e-08 &  $-$4.43e-09 &  $-$2.15e-10 &  $-$1.67e-11 \\
  2.0 & $-$3.25e-02 &  $-$3.49e-04 &  $-$3.92e-06 &  $-$9.71e-09 &   3.34e-10 &   1.01e-11 &  $<$1.0e-12 &  $<$1.0e-12 &  $<$1.0e-12 \\
  3.0 & $-$1.02e-02 &  $-$2.76e-05 &   4.84e-07 &   9.49e-09 &   8.85e-11 &  $<$1.0e-12 &  $<$1.0e-12 &  $<$1.0e-12 &  $<$1.0e-12 \\
  4.0 & $-$2.91e-03 &   2.53e-05 &   4.85e-07 &   4.45e-09 &   3.05e-11 &  $<$1.0e-12 &  $<$1.0e-12 &  $<$1.0e-12 &  $<$1.0e-12 \\
  5.0 &  6.54e-05 &   3.26e-05 &   3.44e-07 &   2.44e-09 &   1.54e-11 &  $<$1.0e-12 &  $<$1.0e-12 &  $<$1.0e-12 &  $<$1.0e-12 \\
 10.0 &  2.50e-03 &   1.66e-05 &   7.47e-08 &   3.09e-10 &   1.28e-12 &  $<$1.0e-12 &  $<$1.0e-12 &  $<$1.0e-12 &  $<$1.0e-12 \\
 20.0 &  1.87e-03 &   5.20e-06 &   1.21e-08 &   2.77e-11 &  $<$1.0e-12 &  $<$1.0e-12 &  $<$1.0e-12 &  $<$1.0e-12 &  $<$1.0e-12 \\
\hline
%\hline
\end{tabular}
\caption{The correlation $(1-{1\over 2}\delta_{n,N/2})S^z(n;T,\Delta)$ for $\Delta=-.1$
for the XXZ spin chain with $N=18$ sites.}
\label{one}
\end{table}

\begin{table}[b]
\begin{tabular}{|c|c|c|c|c|c|c|c|c|c|} 
$ T $ &$n=1$ & $n=2$ & $n=3$ & $n=4$ & $n=5$& $n=6$ & $n=7$& $n=8 $ & $n=9$ \\  
\hline
  0.1 & $-$3.35e-01 &  $-$5.19e-02 &  $-$3.23e-02 &  $-$1.08e-02 &  $-$9.19e-03 &  $-$4.14e-03 &  $-$3.93e-03 &  $-$2.32e-03 &  $-$1.39e-03 \\
  0.2 & $-$3.17e-01 &  $-$4.71e-02 &  $-$2.25e-02 &  $-$6.41e-03 &  $-$3.64e-03 &  $-$1.28e-03 &  $-$7.22e-04 &  $-$3.19e-04 &  $-$1.38e-04 \\
  0.5 & $-$1.97e-01 &  $-$2.13e-02 &  $-$3.09e-03 &  $-$3.32e-04 &  $-$3.77e-05 &  $-$2.78e-06 &  $-$1.70e-07 &   1.06e-08 &   3.20e-09 \\
  1.0 & $-$5.27e-02 &  $-$8.92e-04 &   2.04e-04 &   3.00e-05 &   2.38e-06 &   1.33e-07 &   6.04e-09 &   3.33e-10 &   2.95e-11 \\
  2.0 &  1.40e-02 &   2.18e-03 &   1.49e-04 &   7.98e-06 &   4.09e-07 &   2.16e-08 &   1.16e-09 &   6.30e-11 &   3.37e-12 \\
  3.0 &  2.20e-02 &   1.47e-03 &   6.52e-05 &   2.60e-06 &   1.04e-07 &   4.19e-09 &   1.70e-10 &   6.87e-12 &  $<$1.0e-12 \\
  4.0 &  2.16e-02 &   9.76e-04 &   3.29e-05 &   1.05e-06 &   3.34e-08 &   1.07e-09 &   3.45e-11 &   1.11e-12 &  $<$1.0e-12 \\
  5.0 &  1.98e-02 &   6.82e-04 &   1.87e-05 &   4.92e-07 &   1.30e-08 &   3.47e-10 &   9.22e-12 &  $<$1.0e-12 &  $<$1.0e-12 \\
 10.0 &  1.25e-02 &   1.99e-04 &   2.83e-06 &   4.00e-08 &   5.65e-10 &   8.00e-12 &  $<$1.0e-12 &  $<$1.0e-12 &  $<$1.0e-12 \\
 20.0 &  6.87e-03 &   5.30e-05 &   3.87e-07 &   2.82e-09 &   2.06e-11 &  $<$1.0e-12 &  $<$1.0e-12 &  $<$1.0e-12 &  $<$1.0e-12 \\
\hline
%\hline
\end{tabular}
\caption{The correlation $(1-{1\over 2}\delta_{n,N/2})S^z(n;T,\Delta)$ for $\Delta=-.3$
for the XXZ spin chain with $N=18$ sites.} 
\label{two}
\end{table}

\begin{table}[b]
\begin{tabular}{|c|c|c|c|c|c|c|c|c|c|} 
$ T $ &$n=1$ & $n=2$ & $n=3$ & $n=4$ & $n=5$& $n=6$ & $n=7$& $n=8 $ & $n=9$ \\ 
\hline
  0.1 & $-$5.46e-02 &  $-$2.14e-02 &  $-$3.41e-03 &   5.10e-03 &   8.06e-03 &   8.19e-03 &   7.27e-03 &   6.35e-03 &   2.99e-03 \\
  0.2 &  8.19e-02 &   8.50e-02 &   7.12e-02 &   5.22e-02 &   3.51e-02 &   2.26e-02 &   1.47e-02 &   1.05e-02 &   4.61e-03 \\
  0.5 &  2.02e-01 &   1.36e-01 &   7.38e-02 &   3.62e-02 &   1.72e-02 &   8.18e-03 &   4.03e-03 &   2.23e-03 &   8.66e-04 \\
  1.0 &  2.05e-01 &   8.74e-02 &   3.03e-02 &   9.96e-03 &   3.26e-03 &   1.07e-03 &   3.56e-04 &   1.28e-04 &   3.81e-05 \\
  2.0 &  1.55e-01 &   3.57e-02 &   7.11e-03 &   1.39e-03 &   2.72e-04 &   5.33e-05 &   1.05e-05 &   2.13e-06 &   4.01e-07 \\
  3.0 &  1.19e-01 &   1.83e-02 &   2.55e-03 &   3.51e-04 &   4.83e-05 &   6.66e-06 &   9.19e-07 &   1.29e-07 &   1.75e-08 \\
  4.0 &  9.51e-02 &   1.10e-02 &   1.17e-03 &   1.24e-04 &   1.31e-05 &   1.39e-06 &   1.48e-07 &   1.58e-08 &   1.66e-09 \\
  5.0 &  7.90e-02 &   7.29e-03 &   6.28e-04 &   5.40e-05 &   4.64e-06 &   3.99e-07 &   3.43e-08 &   2.97e-09 &   2.53e-10 \\
 10.0 &  4.23e-02 &   1.94e-03 &   8.53e-05 &   3.76e-06 &   1.66e-07 &   7.30e-09 &   3.22e-10 &   1.42e-11 &  $<$1.0e-12 \\
 20.0 &  2.19e-02 &   4.96e-04 &   1.11e-05 &   2.46e-07 &   5.48e-09 &   1.22e-10 &   2.72e-12 &  $<$1.0e-12 &  $<$1.0e-12 \\
\hline
%\hline
\end{tabular}
\caption{The correlation $(1-{1\over 2}\delta_{n,N/2})S^z(n;T,\Delta)$ for $\Delta=-.9$
for the XXZ spin chain with $N=18$ sites.}
\label{three}
\end{table}

\begin{table}[b]
\begin{tabular}{|c|c|c|c|c|c|c|c|c|c|} 
$ T $ &$n=1$ & $n=2$ & $n=3$ & $n=4$ & $n=5$& $n=6$ & $n=7$& $n=8 $ & $n=9$ \\ 
\hline
  0.1 &  3.30e-01 &   3.19e-01 &   3.02e-01 &   2.83e-01 &   2.64e-01 &   2.47e-01 &   2.33e-01 &   2.25e-01 &   1.11e-01 \\
  0.2 &  3.21e-01 &   2.87e-01 &   2.41e-01 &   1.95e-01 &   1.55e-01 &   1.25e-01 &   1.05e-01 &   9.28e-02 &   4.45e-02 \\
  0.5 &  2.95e-01 &   2.04e-01 &   1.21e-01 &   6.74e-02 &   3.70e-02 &   2.06e-02 &   1.19e-02 &   7.78e-03 &   3.28e-03 \\
  1.0 &  2.50e-01 &   1.14e-01 &   4.41e-02 &   1.63e-02 &   6.04e-03 &   2.24e-03 &   8.46e-04 &   3.50e-04 &   1.14e-04 \\
  2.0 &  1.79e-01 &   4.50e-02 &   9.99e-03 &   2.19e-03 &   4.79e-04 &   1.05e-04 &   2.31e-05 &   5.30e-06 &   1.11e-06 \\
  3.0 &  1.35e-01 &   2.29e-02 &   3.55e-03 &   5.46e-04 &   8.40e-05 &   1.29e-05 &   1.99e-06 &   3.14e-07 &   4.72e-08 \\
  4.0 &  1.07e-01 &   1.37e-02 &   1.62e-03 &   1.92e-04 &   2.27e-05 &   2.68e-06 &   3.17e-07 &   3.80e-08 &   4.43e-09 \\
  5.0 &  8.88e-02 &   9.06e-03 &   8.70e-04 &   8.33e-05 &   7.99e-06 &   7.65e-07 &   7.33e-08 &   7.09e-09 &   6.73e-10 \\
 10.0 &  4.73e-02 &   2.40e-03 &   1.18e-04 &   5.77e-06 &   2.83e-07 &   1.39e-08 &   6.81e-10 &   3.35e-11 &   1.64e-12 \\
 20.0 &  2.44e-02 &   6.13e-04 &   1.52e-05 &   3.77e-07 &   9.33e-09 &   2.31e-10 &   5.73e-12 &  $<$1.0e-12 &  $<$1.0e-12 \\
\hline
%\hline
\end{tabular}
\caption{The correlation $(1-{1\over 2}\delta_{n,N/2})S^z(n;T,\Delta)$ for $\Delta=-1.0$
for the XXZ spin chain with $N=18$ sites.}
\label{four}
\end{table}

\begin{table}[b]
\begin{tabular}{|c|c|c|c|c|c|c|c|c|} 
$ T $ &$n=1$ & $n=2$ & $n=3$ & $n=4$ & $n=5$& $n=6$ & $n=7$& $n=8$  \\ 
\hline
  0.1 & $-$5.62e-02 &  $-$2.25e-02 &  $-$4.06e-03 &   4.90e-03 &   8.37e-03 &   9.12e-03 &   8.91e-03 &   4.37e-03 \\
  0.2 &  7.92e-02 &   8.32e-02 &   7.02e-02 &   5.20e-02 &   3.56e-02 &   2.40e-02 &   1.74e-02 &   7.65e-03 \\
  0.5 &  2.02e-01 &   1.36e-01 &   7.38e-02 &   3.63e-02 &   1.73e-02 &   8.49e-03 &   4.70e-03 &   1.82e-03 \\
  1.0 &  2.05e-01 &   8.74e-02 &   3.03e-02 &   9.96e-03 &   3.26e-03 &   1.08e-03 &   3.90e-04 &   1.16e-04 \\
  2.0 &  1.55e-01 &   3.57e-02 &   7.11e-03 &   1.39e-03 &   2.72e-04 &   5.34e-05 &   1.08e-05 &   2.05e-06 \\
  3.0 &  1.19e-01 &   1.83e-02 &   2.55e-03 &   3.51e-04 &   4.83e-05 &   6.66e-06 &   9.36e-07 &   1.27e-07 \\
  4.0 &  9.51e-02 &   1.10e-02 &   1.17e-03 &   1.24e-04 &   1.31e-05 &   1.39e-06 &   1.49e-07 &   1.56e-08 \\
  5.0 &  7.90e-02 &   7.29e-03 &   6.28e-04 &   5.40e-05 &   4.64e-06 &   3.99e-07 &   3.45e-08 &   2.95e-09 \\
 10.0 &  4.23e-02 &   1.94e-03 &   8.53e-05 &   3.76e-06 &   1.66e-07 &   7.30e-09 &   3.23e-10 &   1.42e-11 \\
 20.0 &  2.19e-02 &   4.96e-04 &   1.11e-05 &   2.46e-07 &   5.48e-09 &   1.22e-10 &   2.72e-12 &  $<$1.0e-12 \\
\hline
%\hline
\end{tabular}
\caption{The correlation $(1-{1\over 2}\delta_{n,N/2})S^z(n;T,\Delta)$ for $\Delta=-.9$
for the XXZ spin chain with $N=16$ sites.}
\label{five}
\end{table}

\begin{table}[b]
\begin{tabular}{|c|c|c|c|c|c|c|c|c|c|} 
$\Delta$ &$n=1$ & $n=2$ & $n=3$ & $n=4$ & $n=5$& $n=6$ & $n=7$& $n=8$ & $n=9$  \\ 
\hline
 $-$0.1 &  4.966 &   3.323 &   2.561 &   2.073 &   1.870 &   1.706 &   1.669 &   1.592 &   ~ \\
 $-$0.2 &  2.432 &   1.643 &   1.275 &   1.037 &   0.923 &   0.840 &   0.811 &   0.774 &   0.767 \\
 $-$0.3 &  1.561 &   1.071 &   0.839 &   0.687 &   0.602 &   0.545 &   0.517 &   0.493 &   0.483 \\
 $-$0.4 &  1.103 &   0.771 &   0.612 &   0.505 &   0.437 &   0.392 &   0.365 &   0.346 &   0.335 \\
 $-$0.5 &  0.807 &   0.578 &   0.464 &   0.388 &   0.334 &   0.297 &   0.272 &   0.253 &   0.243 \\
 $-$0.6 &  0.589 &   0.434 &   0.355 &   0.300 &   0.259 &   0.229 &   0.206 &   0.189 &   0.180 \\
 $-$0.7 &  0.413 &   0.318 &   0.264 &   0.227 &   0.198 &   0.175 &   0.156 &   0.140 &   0.132 \\
 $-$0.8 &  0.265 &   0.215 &   0.184 &   0.161 &   0.142 &   0.126 &   0.112 &   0.099 &   0.094 \\
 $-$0.9 &  0.137 &   0.118 &   0.104 &   0.092 &   0.082 &   0.073 &   0.065 &   0.059 &   0.057 \\
\hline
\end{tabular}
\caption{The values of $T_0(n;\Delta)$ at which the correlation function 
$S^z(n;T_0(n;\Delta),\Delta)$ vanishes for $N=18$}
\label{six}
\end{table}

\begin{table}[b]
\begin{tabular}{|c|c|c|c|} 
$\Delta$ & $T_{0}$ & $\gamma$ & $A$\\
 \hline
$-$0.1 &  1.550 &  0.585  &  5.734  \\
$-$0.2 &  0.745 &  0.547  &  2.690  \\
$-$0.3 &  0.462 &  0.491  &  1.630  \\
$-$0.4 &  0.312 &  0.433  &  1.093  \\
$-$0.5 &  0.216 &  0.374  &  0.764  \\
$-$0.6 &  0.148 &  0.317  &  0.539  \\
$-$0.7 &  0.095 &  0.259  &  0.372  \\
$-$0.8 &  0.054 &  0.205  &  0.241  \\
$-$0.9 &  0.031 &  0.180  &  0.125  \\
\hline
%\hline
\end{tabular}
\caption{The fitting parameters $T_0,\gamma$ and $A$ of (2.10) for 
$\Delta=-.1,\cdots , -.9$}
\label{seven}
\end{table}

{\bf Figure Captions}

Figure 1. A plot of the  exact zeroes $T_0(n;\Delta)$ 
of the $N=18$ system 
compared with the fitting form (2.9). The values $\Delta=-.1,\cdots ,-.9$
are given with $\Delta=-.1$ being the highest.

Figure 2. The temperature $T_0(\Delta)$ plotted as a function of $\Delta.$
\newpage
\centerline{\epsfxsize=6in\epsfbox{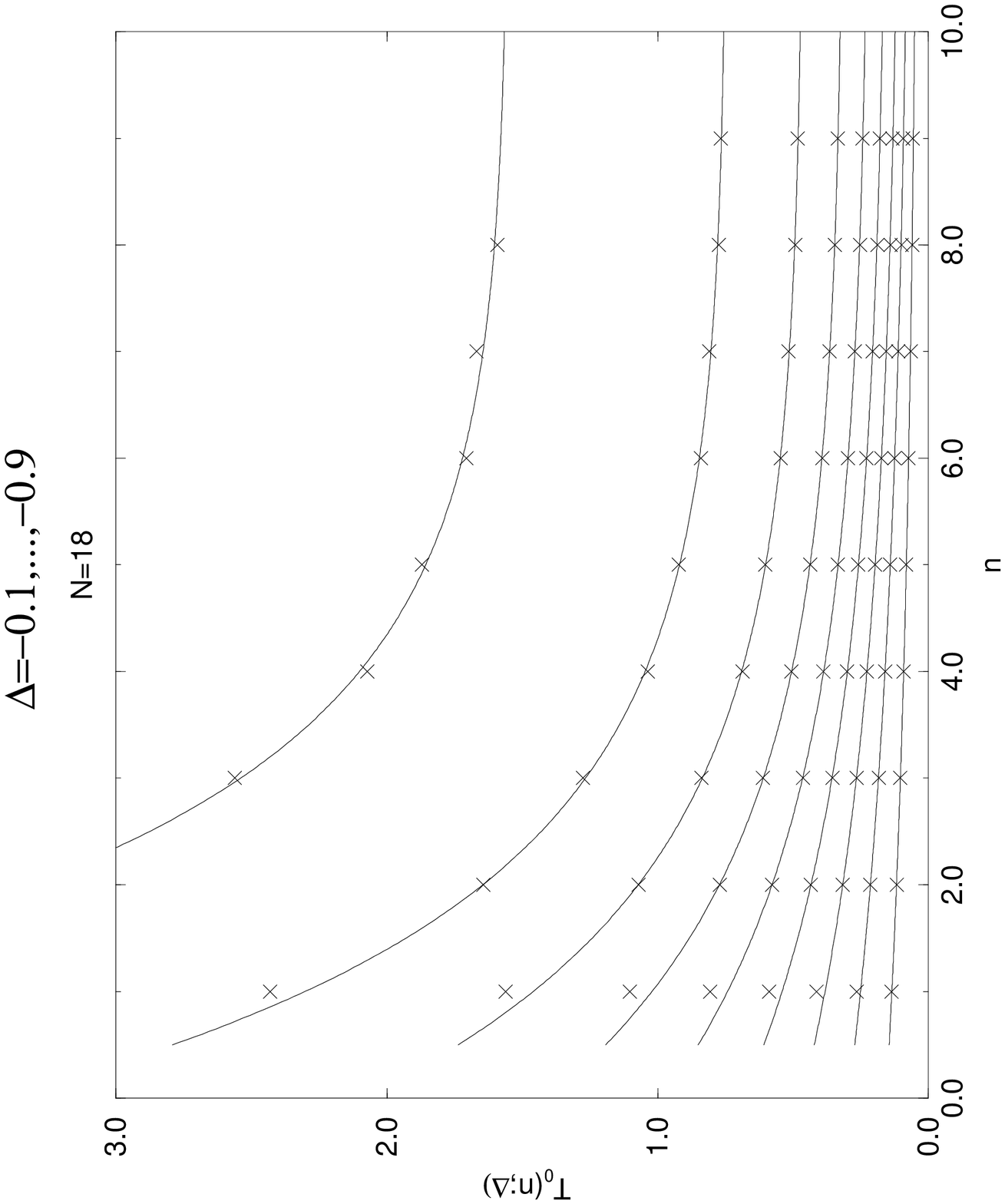}}
\centerline{Fig. 1}
\newpage
\centerline{\epsfxsize=6in\epsfbox{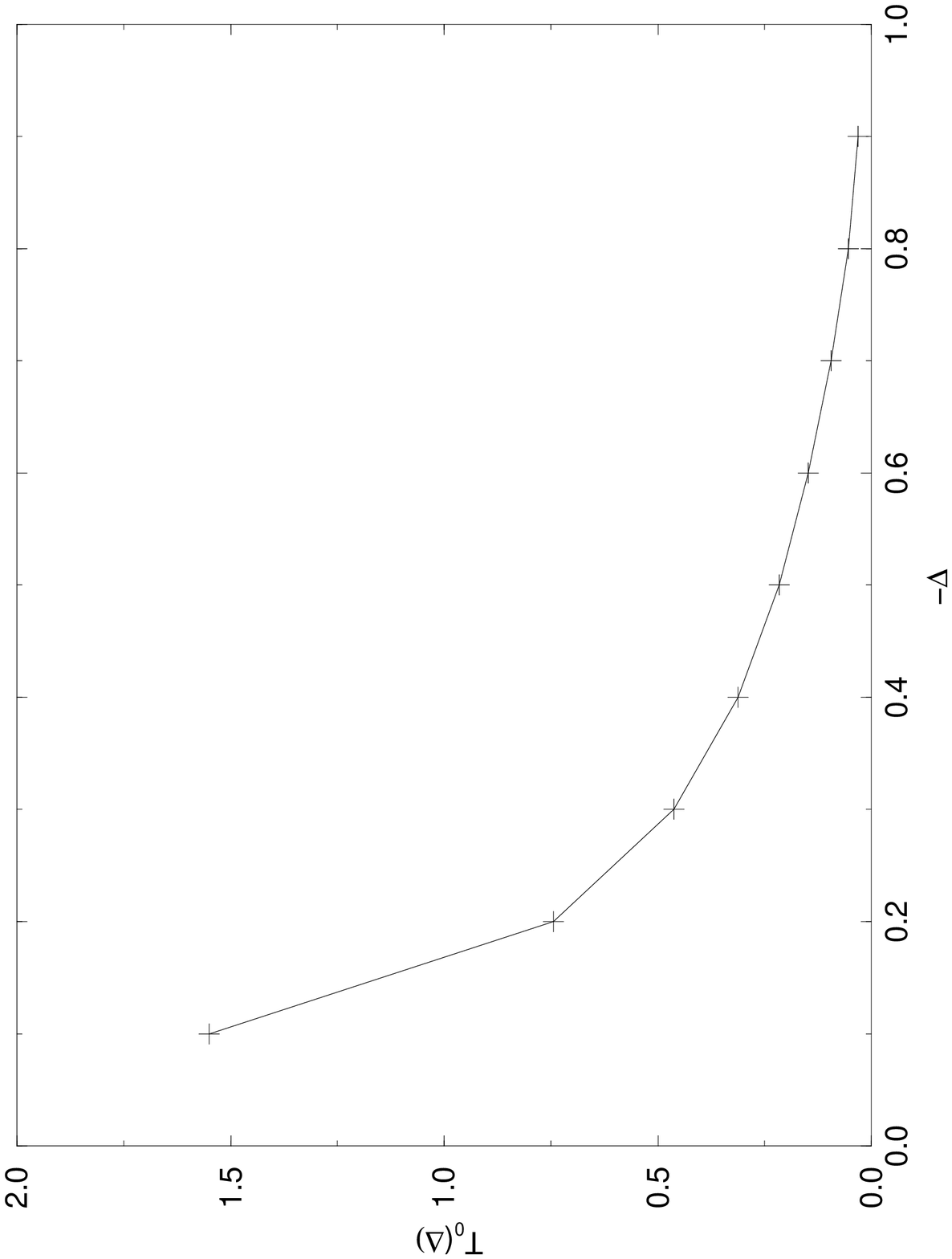}}
\centerline{Fig. 2}


\begin{references}
\bibitem{bet}H.A. Bethe,Zur Theorie der Metalle I. Eigenwerte und
Eigenfunktionen  der linearen Atomkette, Z. Phys. 71 (1931) 205.

\bibitem{lsm}E. Lieb, T. Schultz and D. Mattis, Two soluble models  of
an antiferromagnetic chain, Ann. Phys. 16 (1961) 407.
\bibitem{yya} C.N. Yang and C.P Yang, One-dimensional chain of
anisotropic spin-spin interactions I. Properties of the groundstate
energy per site for an infinite system, Phys. Rev. 150 (1966) 327. 
\bibitem{yyb} C.N. Yang and C.P Yang, One-dimensional chain of
anisotropic spin-spin interactions II. Proof of Bethe's hypothesis for
ground state in finite system, Phys. Rev. 150 (1966) 321. 
\bibitem{yyc} C.N. Yang and C.P Yang, One-dimensional chain of
anisotropic spin-spin interactions III. Applications,
 Phys. Rev. 150 (1966) 321. 
\bibitem{ts} M. Takahashi and M. Suzuki, One-dimensional anisotropic
Heisenberg model at finite temperatures, Prog. Theor. Phys. 48 (1972) 2187.
\bibitem{nm} Th. Niemeijer, Some exact calculations on a chain of
spins 1/2, Physica 36 (1967) 377.
\bibitem{bm}B.M. McCoy, Spin correlation functions of the X-Y model,
Phys. Rev. 173 (1968) 531.

\bibitem{jkm}J.D. Johnson, S. Krinsky and B.M. McCoy, Vertical-arrow
correlation length in the eight-vertex model and the low-lying
excitations of the X-Y-Z hamiltonian, Phys. Rev. A8 (1973) 2526.

\bibitem{lp}A. Luther and I. Peschel, Calculation of critical
exponents in two dimensions from quantum field theory in one
dimension, Phys. Rev. B12 (1975) 3908

\bibitem{fog}H.C. Fogedby, Correlation functions for the
Heisenberg-Ising chain at $T=0$, J. Phys. C 11 (1978) 4767.

\bibitem{perk} J.H.H. Perk, H.W. Capel, G.R.W. Quispel and
F.W. Nijhoff, Finite-temperature correlations for the Ising chain in a
transverse field, Physica A123 (1984) 1.

\bibitem{IK} A.G. Izergin and V.E. Korepin, Correlation functions for
the Heisenberg XXZ antiferromagnet, Comm. Math. Phys. 99 (1985) 271.

\bibitem{lin} H.Q. Lin and D.K. Campbell, Spin-spin correlations in
the one dimensional spin-1/2 antiferromagnetic Heisenberg chain,
J. Appl Phys. 69 (1991) 5947.

\bibitem{jmmn} M. Jimbo, K. Miki, T. Miwa and A. Nakaayashiki,
Correlation functions of the XXZ model for $\Delta<-1,$ Phys. Lett. A
168 (1992) 256.

\bibitem{snw}J. Suzuki, T. Nagao and M. Wadati, Exactly solvable models
and finite size corrections, Int. J. Mod. Phys. B6 (1992) 1119.

\bibitem{klu} A. Kl{\"u}mper, Thermodynamics of the anisotropic spin-1/2
Heisenberg chain and related quantum chains, Z. Phys. B91 (1993) 507.
\bibitem{kwz} A. Kl{\"u}mper, T. Wehner and J. Zittartz, 
Conformal spectrum of the six-vertex model, J. Phys. A. 26 (1993) 2815.

\bibitem{kbi} V.E. Korepin, N.M. Bogoliubov and A.G. Izergin {\it
Quantum Inverse Scattering Method and Correlation Functions} Cambridge
Univ. Press (1993)

\bibitem{jmb}M. Jimbo and T.Miwa, {\it Algebraic Analysis of Solvable
Lattice Models} ,CBMS Regional Conference Series in Mathematics 85,
(Providence, RI, American Mathematical Society, 1994) 

\bibitem{kieu} V.E. Korepin, A.G. Izergin, A.G. Essler and D.B. Uglov,
Correlation function of the spin 1/2 XXX
antiferromagnet. Phys. Lett. A190 (1994) 182.


\bibitem{efik}F.H.L. Essler, H. Frahm, A.G. Izergin and V.E. Korepin,
Determinental representation for correlation functions of spin-1/2 XXX
and XXZ Heisenberg magnets, Comm. Math. Phys. 174 (1995) 191.

\bibitem{jm} M. Jimbo and T. Miwa, Quantum KZ equation with $|q|=1$
and correlation functions of the XXZ model in the gapless regime,
J. Phys. A29 (1996) 2923.
 \bibitem{ss} S. Sachdev, Universal finite temperature crossover functions
of the quantum transition in the Ising spin chain in a transverse
field, Nucl. phys. B464 (1996) 576.

\bibitem{llss} A. Leclaire, F. Lesage, S. Sachdev and H. Saleur,
Finite temperature correlations in the one-dimensional quantum Ising
model, Nucl. Phys. B482[FS] (1996) 579.

\bibitem{ls} S. Lukyanov and A. Zamolodchikov, Exact expectation
values of local fields in quantum sine-Gordon model, Nucl. Phys. B493
(1997) 571.

\bibitem{luk}S. Lukyanov, Low energy effective Hamiltonian for the XXZ
spin chain, cond-mat 9712314

\bibitem{aff} I. Affleck, Exact correlation amplitude for the $S=1/2$
Heisenberg antiferromagnetic chain, cond-mat 9802045.

\bibitem{kss} A. Kuniba, K. Sakai and J. Suzuki, Continued fraction
TBA and functional relations in XXZ model at root of unity,
math.QA/9803056.

\bibitem{car}J.L. Cardy, Conformal invariance and universality in
finite-size scaling, J. Phys. A17 (1984) L385.


\bibitem{affb} I. Affleck, Universal term in the free energy at a
critical point and the conformal anomaly, Phys. Rev. Lett. 56 (1986) 277.
\end{references}
\end{document}